\documentclass[11pt]{article}
 
\usepackage{fullpage,graphicx,amssymb,amsmath,url,xcolor,listings}

\graphicspath{{Fig/},{./}}

\newtheorem{prop}{Proposition}

\def\Jac{\mathrm{Jac}}
\def\dx{\mathrm{d}x}
\def\dy{\mathrm{d}y}
\def\ds{\mathrm{d}s}
\def\dr{\mathrm{d}r}
\def\calX{\mathcal{X}}
\def\bbE{\mathbb{E}}
\def\GL{\mathrm{GL}}
\def\dy{\mathrm{d}y}
\def\KL{\mathrm{KL}}
\def\JS{\mathrm{JS}}
\def\calM{\mathcal{M}}
\def\calC{\mathcal{C}}
\def\dmu{\mathrm{d}\mu}
\def\leftsup#1{{}^{#1}}
\def\st{\ :\ }
\def\bbR{\mathbb{R}}
\def\bbC{\mathbb{C}}
\def\LC{\mathrm{LC}}
\def\diag{\mathrm{diag}}
\def\calP{\mathcal{P}}
\def\Cat{\mathrm{Cat}}
\def\Hellinger{\mathrm{Hellinger}}
\def\calN{\mathcal{N}}
\def\Aut{\mathrm{Aut}}
\def\SPD{\mathrm{SPD}}

\def\Euc{\mathrm{Euc}}
\def\flat{\mathrm{flat}}
\def\dtheta{\mathrm{d}\theta}
\def\deta{\mathrm{d}\eta}

\begin{document}

\title{The analytic dually flat space of the mixture family of two prescribed distinct Cauchy distributions}

\author{Frank Nielsen\footnote{\protect\url{https://franknielsen.github.io/}}\\ Sony Computer Science Laboratories Inc.\\ Tokyo, Japan}

\date{}
\maketitle

\begin{abstract}
A smooth and strictly convex function on an open convex domain induces both 
(1) a Hessian manifold with respect to the standard flat Euclidean connection,
and (2) a dually flat space of information geometry.
We first review these constructions and illustrate how to instantiate them for 
(a) full regular exponential families from their partition functions, 
(b) regular homogeneous cones from their characteristic functions,  
and (c) mixture families from their Shannon negentropy functions.
Although these structures can be explicitly built for many common examples of the first two classes,
the differential entropy of a continuous statistical mixture with distinct prescribed density components sharing the same support is hitherto not known in closed form, hence forcing implementations  of mixture family manifolds  in practice using Monte Carlo sampling. 
In this work, we report a notable exception: 
The family of mixtures defined as the convex combination of two prescribed and distinct Cauchy distributions.
As a byproduct, we report closed-form formula for the Jensen-Shannon divergence between two mixtures of two prescribed Cauchy components.
\end{abstract}


\section{Introduction and motivation}  

Information geometry~\cite{IG-2016} studies the geometric structures of a family $\calP=\{p_\theta(x)\}_{\theta\in\Theta}$ of parametric probability distributions called the statistical model.
The dimension $D$ of the parameter space $\Theta\subset\bbR^D$ denotes the order of the model (e.g., $D=1$ for the family of exponential distributions, $D=2$ for the family of univariate normal distributions, etc.).
Information geometry relies on the core concept of affine connections in differential geometry~\cite{IntroRieGeo-2012}.
Amari pioneered this field and elicited in particular the so-called dualistic $\pm\alpha$-structure~\cite{IG-2016} 
$(\calP,g_F,\nabla^{-\alpha},\nabla^\alpha)$ of $\calP$  by using a family of affine connections, called the $\alpha$-connections $\nabla^\alpha$ for $\alpha\in\bbR$. 
An $\alpha$-connection~\cite{IG-2016} defines $\nabla^\alpha$-geodesics on $\calP$, and is specified according to its corresponding  
$D^3$ Christoffel symbols $\Gamma^\alpha_{ki,j}$ (functions) as follows:
$$
\Gamma^\alpha_{ki,j}(\theta)=E_{p_\theta}\left[\left(\partial_{k} \partial_{i} l_\theta(x)+\frac{1-\alpha}{2} \partial_{k} l_\theta(x) \partial_{i} l_\theta(x)\right) \partial_{j} l_\theta(x)\right],
$$ 
where $l_\theta(x)=\log p_\theta(x)$ denotes the log-likelihood function, and $\partial_i$ is the notational shortcut for $\frac{\partial}{\partial\theta_i}$ for $i\in\{1,\ldots,D\}$. 
The Riemannian metric tensor $g_F$ is the Fisher information metric expressed in the $\theta$-coordinate system using the Fisher information matrix as follows:
$$
[g_F]_\theta = E_{p_\theta}\left[\nabla_\theta l_\theta(x)  (\nabla_\theta l_\theta(x))^\top \right].
$$
For any $\alpha\in\bbR$, the connections $\nabla^{-\alpha}$ and $\nabla^\alpha$ are proven dual with respect to the Fisher information metric $g_F$ since their mid connection $\frac{\nabla^{-\alpha}+\nabla^\alpha}{2}$ corresponds to the Levi-Civita metric connection $\leftsup{g}\nabla$~\cite{IntroRieGeo-2012}. 
The fundamental theorem of Riemannian geometry~\cite{IntroRieGeo-2012} states that the Levi-Civita metric the unique torsion-free metric-compatible affine connection.

Two common types of families of probability distributions are considered in information geometry: 
The exponential families~\cite{Barndorff-2014} and the mixture families~\cite{IG-2016,MCIG-2019}.
The $\pm1$-structures (i.e., $\alpha$-structure for $\alpha=\pm 1$) of the exponential families and  mixture families are said dually flat (to be detailed in \ref{sec:geostruct}).
Dually flat spaces have also been called Bregman manifolds~\cite{geodesictrianglesDFS-2021} as they can be realized from either a smooth and strictly convex functions (Bregman generators) or equivalently from their corresponding Bregman divergences via the information geometry structure derived from divergences~\cite{amari2010information}.

However, there is a significant difference when considering the Bregman generators induced by exponential families from the Bregman generators induced by mixture families: 
While the Bregman generators of exponential families (i.e., cumulant functions)  are always real analytic  and available in closed form for many common exponential families (i.e., multivariate normal family or Beta family), 
the Bregman generators of mixture families (Shannon negentropy of mixtures) can be non-analytic~\cite{Watanabe-2004} (e.g., Shannon negentropy of a mixture of two prescribed and distinct Gaussian components).
Let us notice that the family of categorical distributions can be both interpreted as a discrete mixture family and a discrete exponential family~\cite{IG-2016} with closed-form Bregman generators being convex conjugate of each other~\cite{IG-2016}.

In this work, we present a mixture family $\calC=\{(1-\theta)p_{l_1,s_1}+\theta p_{l_2,s_2}\ :\ \theta\in(0,1)\}$  of two distinct Cauchy distributions $p_{l_1,s_1}$ and $p_{l_2,s_2}\not=p_{l_1,s_1}$ (order $D=1$) which is analytic, and report the convex conjugate Bregman functions and dual parameterizations in closed-form.

The paper is organized as follows: 
In section~\ref{sec:geostruct}, we recall the two usual differential-geometric constructions obtained from a convex function in an open convex domain of $\bbR^d$:
Namely,  
(1) the Hessian manifold in \S\ref{sec:Hessian}, 
and (2) the dually flat space in \S\ref{sec:Bregman}.
Furthermore, we link those structures to Amari's $\pm1$-structures of exponential families and mixture families.
We then illustrate these constructions in section~\ref{sec:examples} for 
(a) the full regular exponential families~\cite{Barndorff-2014} (\S\ref{sec:EF}), 
(b) the homogeneous regular cones~\cite{BarrierConeFormulaCharacteristic-1996} (\S\ref{sec:cone})
, and (c) the mixture families~\cite{MCIG-2019} (\S\ref{sec:MF}).
Our main contribution is presented in section~\ref{sec:CauchyMix} where we report in closed-form all the necessary formula required to explicitly implement the dually flat space of statistical mixtures of two distinct Cauchy components.
Appendix~\ref{sec:maxima} provides a computational notebook using the open source symbolic computing software {\sc Maxima}~\cite{calvo2018scientific}.

\section{Differential-geometric structures induced by smooth convex functions}\label{sec:geostruct}

\subsection{Hessian manifolds and Bregman manifolds}\label{sec:Hessian}

Consider the $D$-dimensional Euclidean space $\mathbb{E}^d$ as an affine space equipped with the Cartesian coordinate system $x(\cdot)$.
We can view $\mathbb{E}^d$ as a flat manifold $(\mathbb{E}^d,\leftsup{\Euc}\nabla)$ where $\leftsup{\Euc}\nabla$ denotes the standard flat connection of $\mathbb{E}^d$~\cite{Shima-2007} (Chapter~1) such that 
$\leftsup{\Euc}\nabla_{\frac{\partial}{\partial x^i} } \frac{\partial}{\partial x^j}=0$.

In general, the Hessian operator~\cite{IntroRieGeo-2012} applied to a function $F$ on a manifold $M$ is defined according to a connection $\nabla$ (specified by using its Christoffel symbols $\Gamma_{i j}^{k}$):
$$
\nabla^{2} F\left(\partial_{x^{i}}, \partial_{x^{j}}\right)=\frac{\partial^{2} F}{\partial x^{i} \partial x^{j}}-\Gamma_{i j}^{k} \frac{\partial F}{\partial x^{k}}.
$$

A connection is said flat when there exists a coordinate system such that all Christoffel symbols vanish: $\Gamma_{i j}^{k}(x)=0$.
On a flat manifold $(M,\leftsup{\flat}\nabla)$, we thus have the Hessian operator rewritten as:
$$
\nabla^{2} F\left(\partial_{x^{i}}, \partial_{x^{j}}\right)=\frac{\partial^{2} F}{\partial x^{i} \partial x^{j}}.
$$
In particular, this holds on the Euclidean manifold $(\mathbb{E}^d,\leftsup{\Euc}\nabla)$.

A Riemannian metric $g$ on a flat manifold $(M,\nabla)$ is called a Hessian metric~\cite{Shima-2007} (Chapter~2) if there exists 
a local coordinate system $x$ and a potential function $F(x)$ such that $g=[g_{ij}]_{ij}$ with $g_{ij}$ which can be expressed as
$$
g_{ij}(x)=\frac{\partial^2}{\partial x_i\partial x_j} F(x).
$$

When $\nabla=\leftsup{\Euc}\nabla$, we further say that $g$ is a Bregman metric (thus a special type of Hessian metric).
For example, consider a smooth and strictly convex function $F(\theta)$ defined on an open convex domain $\Theta\subset\bbR^D$.
Then
$$
g=\sum_{i=1}^D\sum_{j=1}^D \frac{\partial^2}{\partial \theta_i\partial \theta_j} F(\theta) \dtheta^i\otimes \dtheta^j
$$ 
is a Bregman metric~\cite{gomes2019geometry}.

For a Hessian metric $[g_{h_F}(\theta)]$ with   $h_F(\theta):=\frac{\partial^{2} F}{\partial \theta^{i} \partial \theta^{j}}$, we can associate the corresponding Riemannian distance $\rho_{h_F}(p_1,p_2)$ between any two point points $p_1$, $p_2$ on the Riemannian manifold $(M,g_{h_F})$.
When $h_F$ is the Fisher information matrix, this distance is called the Rao's distance~\cite{atkinson1981rao} or the Fisher-Rao distance~\cite{pinele2020fisher}. 

In the particular interesting case of separable Bregman generators~\cite{gomes2019geometry} with
smooth strictly convex generator $F(\theta)=\sum_{i=1}^D F_i(\theta^i)$  (where the  univariate functions $F_i$'s are scalar Bregman generators), we get the following diagonal Hessian matrix:
$$
h_F(\theta)=\diag(F_1''(\theta^1),\ldots,F_D''(\theta^D)),
$$
and the corresponding Riemannian distance $g_{h_F}$, called the Riemannian Bregman distance in~\cite{gomes2019geometry}, can be computed using the following formula:
\begin{equation}\label{eq:RiemannBregmanDist}
\rho_F(\theta_1,\theta_2)= \sqrt{\sum_{i=1}^D \left(h_i(\theta_1^i)-h_i(\theta_2^i)\right)^2},
\end{equation}
where the functions $h_i$'s are the antiderivatives of $\sqrt{F_i''(u)}$:
$$
h_i(\theta)=\int^\theta \sqrt{F_i''(u)} \mathrm{d}u.
$$
When $F_i(\theta)=\frac{1}{2}\theta^2$ for $i\in\{1,\ldots, D\}$, we have $h_i(\theta)=\theta$, and we recover the usual Euclidean distance formula expressed in the Cartesian coordinate system.
In general, the formula of Eq.~\ref{eq:RiemannBregmanDist}  is the Euclidean distance expressed using the $h(\theta)=(h_1(\theta^1),\ldots, h_D(\theta^D))$-coordinate system.
Indeed, recall that a Riemannian metric tensor $g$ is the Euclidean metric~\cite{IntroRieGeo-2012} if there exists a coordinate system $h$ such that $[g]_h=I$, the identity matrix of dimension $D\times D$. 
The Euclidean metric expressed in the Cartesian coordinate system $\lambda$ is $[g]_\lambda=I$, the identity matrix. 

Notice that the Euclidean distance between two point $p_1$ and $p_2$ of the Euclidean plane $\bbE^2$   is expressed 
in the polar coordinate $(r=\sqrt{x^2+y^2},\theta=\arctan\frac{y}{x})$ (with inverse transformation $(x=r\cos\theta,y=r\sin\theta)$) as
$$
\rho_\Euc(p_1,p_2)= \sqrt{r_1^2+r_2^2-2r_1r_2\cos(\theta_2-\theta_1)}.
$$

The Riemannian metric of the Euclidean metric expressed in the Cartesian coordinate system $\lambda=(x,y)$ is
$\ds^2=\dx^2+\dy^2$, or $\ds^2=[\dx\ \dy]^\top\, G_\lambda(\lambda)\, [\dx\ \dy]$ with $G_\lambda(\lambda)=\diag(1,1)$.
The Euclidean metric tensor can be expressed in any new $\eta$-coordinate system using the following covariant rule:

$$
G_\eta(\eta)=\left[\frac{\partial\lambda_i}{\partial\eta_j}\right]^\top_{ij}\times G_\lambda(\lambda(\eta))\times \left[\frac{\partial\lambda_i}{\partial\eta_j}\right]_{ij},
$$
where $\left[\frac{\partial\lambda_j}{\partial\eta_j}\right]_{ij}$ is the invertible Jacobian matrix of the transformation.
We have 
$$
\left[\frac{\partial\lambda_j}{\partial\eta_j}\right]_{ij}=\left[\frac{\partial (x=r\cos\theta,y=r\sin\theta)}{\partial(r,\theta)}\right]_{ij}=\left[\begin{array}{cc} \cos\theta & -r\sin\theta\\ \sin\theta & r\cos\theta\end{array}\right].$$

Thus it follows that
$$
G_\eta(\eta)= 
\left[\begin{array}{cc} \cos\theta & -r\sin\theta\\ \sin\theta & r\cos\theta\end{array}\right]^\top\times I\times 
\left[\begin{array}{cc} \cos\theta & -r\sin\theta\cr \sin\theta & r\cos\theta \end{array}\right]
 =
\left[\begin{array}{cc} 1 & 0 \cr 0 & r^2\end{array}\right]=\diag(1,r^2),
$$
using the identity $\sin^2\theta+\cos^2\theta=1$.

Hence, the Euclidean metric expressed in the polar coordinate system is 
$$
\ds^2_\eta=[\dr\ \dtheta]^\top\times \diag(1,r^2)\times [\dr\ \dtheta]=\dr^2+r^2\dtheta^2.
$$

The Poincar\'e metric on the plane is defined in the Cartesian coordinate system $\lambda $ by the metric tensor $[g_P]_\lambda=\frac{1}{y^2}\, I$.
This Poincar\'e metric cannot be expressed as  $\Jac_{h(x,y)}^\top \times I \times \Jac_{h(x,y)}$ for an invertible coordinate transformation $h(x,y)$. That is, it is not the Euclidean metric in disguise, but the hyperbolic metric.

\subsection{Bregman manifolds: Dually flat spaces}\label{sec:Bregman}

A dually flat space~\cite{Shima-2007,IG-2016} (also called a Bregman manifold in~\cite{geodesictrianglesDFS-2021}) can be built from any strictly  convex and smooth function $F(\theta)$ (with open convex domain $\mathrm{dom}(F)=\Theta\not=\emptyset$) of Legendre-type~\cite{LegendreType-1967}. The Legendre-Fenchel transformation of $(\Theta, F(\theta))$ yields a dual Legendre-type potential function $(H,F^*(\eta))$ (with open convex domain $\mathrm{dom}(F^*)=H$) where 
$$
F^*(\eta):=\sup_{\theta\in\Theta} \{\theta^\top\eta-F(\theta)\}.
$$ 

\begin{figure}
\centering
\includegraphics[width=0.6\textwidth]{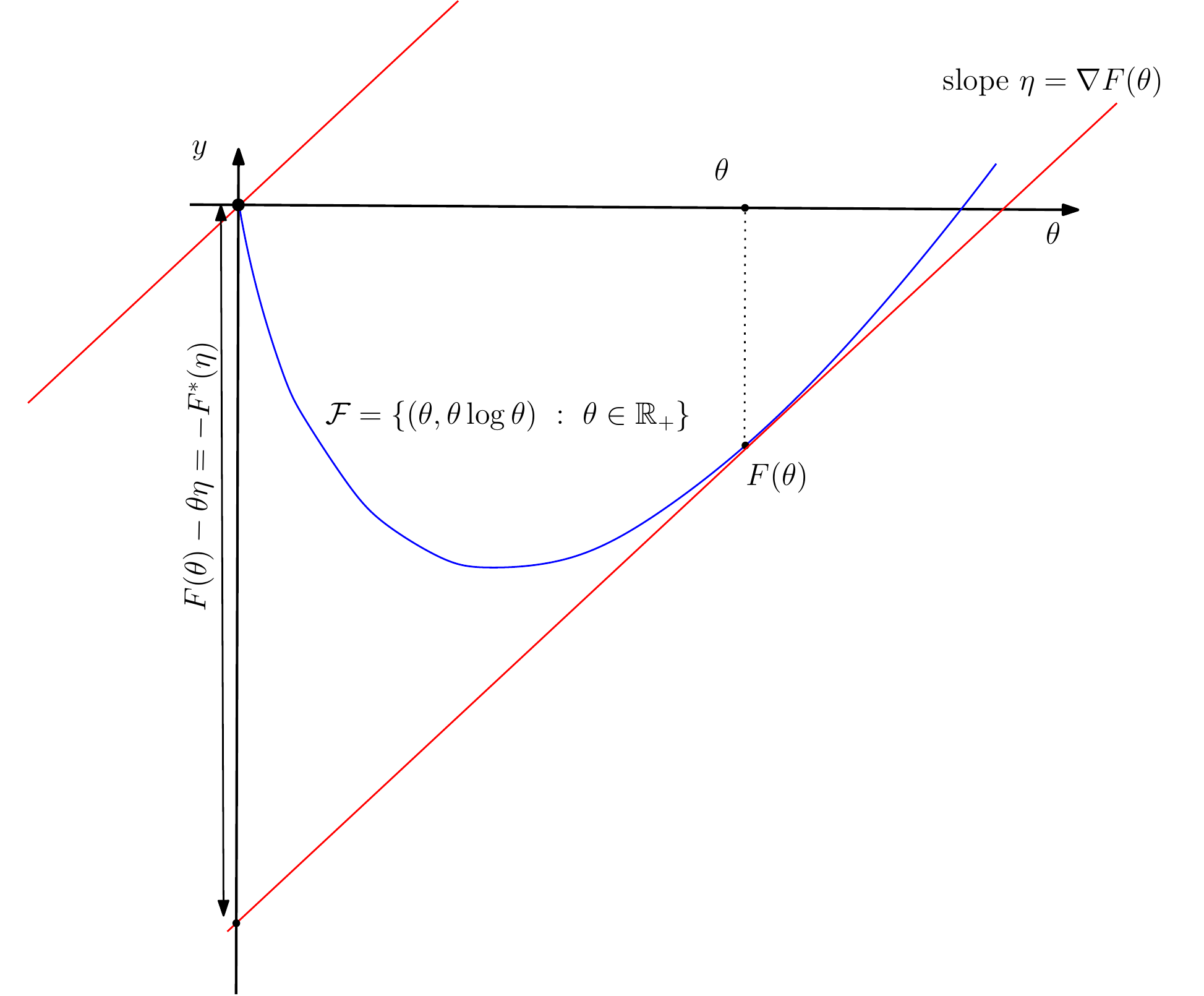}

\caption{Reading geometrically the Legendre-Fenchel transformation as the negative of the $y-$intercept of the the unique tangent line to the graph of $F(\theta)$ which has slope $\eta$.}
\label{fig:Legendre}
\end{figure}

Figure~\ref{fig:Legendre} geometrically interprets the Legendre-Fenchel transform as the negative of the $y$-intercept of the unique tangent line  to the graph of $\mathcal{F}=(\theta,F(\theta))\ :\ \theta\in\Theta\}$ which has slope $\eta$.
 
The Legendre-Fenchel transformation on Legendre-type functions is involutive (i.e., $(F^*)^*=F$ by the Fenchel-Moreau theorem) and induces two dual coordinate systems: 
$$
\eta(\theta)=\nabla_\theta F(\theta),
$$ 
and 
$$
\theta(\eta)=\nabla_\eta F^*(\eta).
$$ 
Thus the gradients of convex conjugates are inverse functions of each other: $\nabla F^*=(\nabla F)^{-1}$ and $\nabla F=(\nabla F^*)^{-1}$.

The Bregman manifold is equipped with a divergence: 
$$
B_F(\theta_1:\theta_2):=F(\theta_1)-F(\theta_2)-(\theta_1-\theta_2)^\top\nabla F(\theta_2),
$$ 
for the  Bregman generator $F(\theta)$ called the Bregman divergence~\cite{bregman1967relaxation}, and we have 
$B_F(\theta_1:\theta_2)=B_{F^*}(\eta_2:\eta_1)$.
We can also express equivalently the dual Bregman divergences using mixed parameterizations  with the Fenchel-Young divergences~\cite{Blondel-2020,geodesictrianglesDFS-2021}:
$$
Y_F(\theta_1:\eta_2):=F(\theta_1)+F^*(\eta_2)-\theta_1^\top\eta_2.
$$ 
Thus we can express the divergence using either primal, dual, or mixed coordinate systems as follows: 
$$
B_F(\theta_1:\theta_2)=Y_F(\theta_1:\eta_2)=Y_{F^*}(\eta_2:\theta_1)=B_{F^*}(\eta_2:\eta_1).
$$

A Riemannian Hessian metric tensor~\cite{Shima-2007} $\leftsup{F}g$ can be defined in the $\theta$-coordinate system by 
$$
[\leftsup{F}g]_\theta:=\nabla^2_\theta F(\theta),
$$ 
with dual metric tensor $\leftsup{F^*}g$ expressed in the $\eta$-coordinate system by 
$$
[\leftsup{F^*}g]_\eta:=\nabla^2_\eta F^*(\eta).
$$
Let $\theta=(\theta^1,\ldots,\theta^D)$ denote the contravariant coordinates and $\eta=(\eta_1,\ldots,\eta_D)$ its equivalent covariant coordinates.
Let $\partial_i:=\frac{\partial}{\partial\theta^i}$ define the primal natural basis $E:=\{e_i=\partial_i\}$, and 
let $\partial^i:=\frac{\partial}{\partial\eta_i}$ define the dual natural basis $E^*:=\{{e^*}^i=\partial^i\}$.
We have 
$$
\leftsup{F}g(e_i,e_j)=\partial_i\partial_j F(\theta),\quad \leftsup{F^*}g({e^*}^i,{e^*}^j)=\partial^i\partial^j F^*(\eta).
$$

The Crouzeix identity~\cite{Crouzeix-1977} holds (i.e., $\nabla^2_\theta F(\theta)\nabla^2_\eta F^*(\eta)=I$, the identity matrix) meaning
that the basis $E$ and $E^*$ are reciprocal~\cite{IG-2016,EIG-2020}: $g(e_i,{e^*}^j)=\delta_{i}^j$, where 
$\delta_{i}^j$ is the Kr\"onecker symbol: $\delta_{i}^j=0$ if $j\not=i$, and $\delta_{i}^j=1$ iff. $i=j$.

The Riemannian metric tensor can thus be expressed equivalently as
\begin{eqnarray*}
\leftsup{F}g &=& \frac{\partial}{\partial\theta^i}\frac{\partial}{\partial\theta^j} F(\theta)\, \dtheta^i \otimes\dtheta^j,\\
&=& \frac{\partial}{\partial\eta_i}\frac{\partial}{\partial\eta_j} F^*(\eta)\, \deta^i \otimes \deta^j,\\
&=& \dtheta^i \otimes \deta^j,
\end{eqnarray*}
where $\otimes$ denotes the tensor product~\cite{IntroRieGeo-2012}.

A Bregman manifold has been called a dually flat space in information geometry~\cite{IG-2016,EIG-2020} because the dual potential functions $F(\theta)$ and $F^*(\eta)$ induce two affine connections, denoted by $\leftsup{F}\nabla$ and $\leftsup{F^*}\nabla$, which are flat because their corresponding Riemann-Christoffel symbols $\leftsup{F}\Gamma_{ij}^k$  characterizing $\leftsup{F}\nabla$ vanish  in the $\theta$-coordinate system (i.e.,  $\leftsup{F}\Gamma_{ij}^k(\theta)=0$ and $\theta(\cdot)$ is called a $\leftsup{F}\nabla$-coordinate system) and the Riemann-Christoffel symbols $\leftsup{F^*}\Gamma_{ij}^k$  characterizing $\leftsup{F^*}\nabla$ vanish in the $\eta$-coordinate system 
(i.e., $\leftsup{F^*}\Gamma_{ij}^k(\eta)=0$ and $\eta(\cdot)$ is called a $\leftsup{F^*}\nabla$-coordinate system).
Furthermore, the two (torsion free) affine connections $\leftsup{F}\nabla$ and $\leftsup{F^*}\nabla$ are dual with respect to the metric tensor~\cite{IG-2016,EIG-2020} $\leftsup{F}g$ so that we have the mid-connection which coincides with the Levi-Civita metric connection: 
$$
\frac{\leftsup{F}\nabla+\leftsup{F^*}\nabla}{2}=\leftsup{\LC}\nabla,
$$ 
where $\leftsup{\LC}\nabla$ denote the Levi-Civita connection induced by the Hessian metric $\leftsup{F}g$.

Two common examples of dually flat spaces of statistical models are the exponential family manifolds~\cite{IG-2016,EIG-2020} built from regular exponential families~\cite{Barndorff-2014} by setting the Bregman generators to the cumulant functions of the family, and the mixture family manifolds induced by the negentropy of a statistical mixture with prescribed linearly independent component distributions~\cite{IG-2016,EIG-2020}.
The family of categorical distributions (also called multinoulli distributions) are both an exponential family and a mixture family. 
It is interesting to notice that the cumulant functions of regular exponential families are always analytic ($C^\omega$, see~\cite{Barndorff-2014}), i.e., $F(\theta)$ admitting locally a  converging Taylor series at any $\theta\in\Theta$) but the negentropy of a mixture may not be analytic (e.g., negentropy of a mixture of two normal distributions~\cite{Watanabe-2004}).

To use the toolbox of geometric algorithms on Bregman manifolds (e.g.,~\cite{Banerjee-2005,BVD-2010}), one needs the generators $F$ and $F^*$ and their gradient $\nabla F$ and $\nabla F^*$ in closed-form. This may not always be possible~\cite{MCIG-2019} either:
\begin{itemize}
\item  because it is not {\em computable} using elementary functions (e.g., the cumulant function of a polynomial exponential family) or  the negentropy of a Gaussian mixture~\cite{Watanabe-2004} (definite integral of a log-sum-exp term), or 

\item because it is {\em computationally intractable} (e.g., the cumulant function of a discrete exponential family in Boltzman machines~\cite{IG-2016})
\end{itemize}

Eguchi~\cite{eguchi1992geometry} described the following method to build a dual information-geometric structure from a smooth parameter divergence $D(\cdot:\cdot)$ which meets the following requirements.
\begin{enumerate}
\item $D(\theta:\theta')\geq 0$ for all $\theta,\theta'$ with equality iff $\theta=\theta'$.

\item ${\partial_i}\left.D(\theta:\theta')\right|_{\theta'=\theta}={\partial_j'}\left.D(\theta:\theta')\right|_{\theta'=\theta}=0$ for all $i,j$, where $\partial_l:=\frac{\partial}{\partial\theta_l}$ and $\partial_l':=\frac{\partial}{\partial\theta_l'}$.

\item  $-\left[{\partial_i}{\partial_j'}\left.D(\theta:\theta')\right|_{\theta'=\theta}\right]_{ij}$ is  a positive-definite matrix.
\end{enumerate}

The construction, called divergence information geometry~\cite{amari2010information}, proceeds as follows:
\begin{eqnarray*}
g_{ij}(\theta)&=&-{\partial_i}{\partial_j'}\left.D(\theta:\theta')\right|_{\theta'=\theta},\\
\Gamma_{ij,k}(\theta)&=&-{\partial_i}{\partial_j}{\partial_k'}\left.D(\theta:\theta')\right|_{\theta'=\theta},\\
\Gamma_{ij,k}^*(\theta) &=& -{\partial_k}{\partial_i'}{\partial_j'}\left.D(\theta:\theta')\right|_{\theta'=\theta}.
\end{eqnarray*}
It can be shown that the connections $\nabla$ and $\nabla^*$ induced respectively by $\Gamma_{ij,k}$ and $\Gamma_{ij,k}^*$ are torsion-free and dual. 

In practice, many explicit dually flat space constructions have been reported for exponential families (e.g.,~\cite{zhang2007information,IG-Dirichlet-2008,IG-MVN-Malago-2015}) but to the best of the author's knowledge none so far for continuous mixture families with component distributions sharing the real-line support.

We report a first exception: The explicit construction of a dually flat space of the family of statistical mixtures with two prescribed  and distinct Cauchy distributions. 
That is, we report in closed-form the Bregman generator $F(\theta)$, the dual parameter $\eta=F'(\theta)$, the dual Bregman generator $F^*(\eta)$ and its derivative $(F^*)'(\eta)=\theta$, and the dual Bregman divergences $B_F(\theta_1:\theta_2)=B_{F^*}(\eta_2:\eta_1)$ which amount to the Kullback-Leibler divergence between the corresponding Cauchy mixtures.
We check these (large) formula using symbolic calculations, and to fix ideas instantiate these formula for the special case of a mixture family which is obtained as the convex combination of a standard Cauchy density (location parameter $0$ and scale parameter $1$) with the Cauchy density of location parameter $1$ and scale parameter $1$.

\section{Some illustrating examples}\label{sec:examples}

\subsection{Exponential family manifolds}\label{sec:EF}

\subsubsection{Natural exponential family}\label{sec:natexpfam}

A natural exponential family~\cite{Barndorff-2014} $\mathcal{E}=\{P_\theta\}$ in a probability space $(\calX,\Sigma,\mu)$ is a set of parametric probability measures $P_\theta$ all dominating by $\mu$ (on support $\calX$) with Radon-Nikodym densities $p_\theta=\frac{\mathrm{d}P_\theta}{\dmu}$ which can be expressed canonically as
$$
p_\theta(x)=\exp(\sum_{i=1}^D \theta^i x^i-F(\theta)),
$$
where $F(\theta)=\log\int \exp(\theta x)\dmu(x)$ is called the cumulant function.
It can be shown that $Z(\theta)=\exp(F(\theta))$ is logarithmically strictly convex~\cite{Barndorff-2014}, and thus $F(\theta)$ is strictly convex. Moreover, $F(\theta)$ is real analytic on the natural parameter space $\Theta=\{\theta\: \: F(\theta)<\infty\}$.
Thus on the domain $\Theta\subset\bbR^D$, $F(\theta)$ induces a Bregman manifold and a dually flat space structures.
The family of categorical distributions form a discrete exponential family~\cite{IG-2016}.

More generally, the concept of an exponential family can be generalized in non-statistical contexts~\cite{NonProbaEF-2012}.

\subsubsection{Fisher-Rao manifold of the categorical distributions}\label{sec:catexpfam}
Consider the family of categorical distributions 
$\calP=\{p_\theta\st \theta\in\Theta\}$ on the probability space $(\Omega_d,2^{\Omega_d},\mu_\#)$, 
where $\mu_\#$ denotes the counting measure on the sample space $\Omega_d=\{\omega_1,\ldots,\omega_d)$.
When $d=2$, the categorical distributions are called the Bernoulli distributions,
and when $d>2$ they are sometimes termed the multinoulli distributions.

The density of a random variable $X\sim \Cat(q_1,\ldots,q_d)$ following a categorical distribution $p_\theta(x)$ is
$$
p_\theta(x)=\prod_{i=1}^d q_i^{x_i},\quad \forall i\in\{1,\ldots, d\}, x_i\in\{0,1\}, \sum_{i=1}^d x_i=1,
$$
where $q_i=\Pr(X=\omega_i)$.
The parameter space $\Theta$ is the $(d-1)$ dimensional open standard simplex $\Delta_d^\circ$.

The Fisher information matrix (FIM) of the categorical distributions is
\begin{eqnarray*}
I_\theta(\theta) &=& E_{p_\theta}\left[ \nabla \log p_\theta(x) \left( \nabla \log p_\theta(x)\right)^\top \right],\\
&=& \left[ E_{p_\theta}\left[\frac{\partial}{\partial\theta_i}\log p_\theta(x) \frac{\partial}{\partial\theta_j}\log p_\theta(x)\right] \right]_{ij},\\
&=& \left[ E_{p_\theta}\left[ \frac{x_ix_j}{\theta_i\theta_j}\right] \right]_{ij}.
\end{eqnarray*}

\begin{itemize}
\item When $i=j$, we have
$$
[I_\theta(\theta)]_{ii} =\frac{1}{\theta_i^2} E_{p_\theta}[x_i^2]=\frac{1}{\theta_i},
$$
since $E_{p_\theta}[x_i^2]=x_i^2 q_i=q_i=\frac{1}{\theta_i}$.
\item When $i\not=j$, we have
$$
[I_\theta(\theta)]_{ij} =\frac{1}{\theta_i\theta_j} E_{p_\theta}[x_ix_j]=0,
$$
since $x_ix_j=0$ for $i\not=j$ (because we have $\forall i\in\{1,\ldots, d\}, x_i\in\{0,1\}$ and $\sum_{i=1}^d x_i=1$).
\end{itemize}

Thus it follows that the Fisher information matrix of the categorical distributions is the diagonal matrix:
$$
I_\theta(\theta)=\diag\left(\frac{1}{\theta_1},\ldots,\frac{1}{\theta_d}\right).
$$

For any smooth invertible mapping $\eta(\theta)$ with invertible Jacobian matrix $\left[\frac{\partial\theta_j}{\partial\eta_j}\right]_{ij}$, we have the following covariant rule of the FIM:
$$
I_\eta(\eta)=\left[\frac{\partial\theta_i}{\partial\eta_j}\right]^\top_{ij}\times I_\theta(\theta(\eta))\times \left[\frac{\partial\theta_i}{\partial\eta_j}\right]_{ij}.
$$

Thus by making a smooth change of variable $\theta\mapsto (\eta_1(\theta),\ldots,\eta_d(\theta))$ with $\eta_i(\theta)=2\sqrt{\theta_i}$ (with $\eta_i(\theta)=2\sqrt{\theta_i}$), we get the following Jacobian matrix 
$$
\left[\frac{\partial\theta_i}{\partial\eta_j}\right]^\top_{ij}
=\diag(\frac{1}{2}\eta_1(\theta),\ldots,\frac{1}{2}\eta_d(\theta))=\diag(\sqrt{\theta_1},\ldots,\sqrt{\theta_d}).
$$
Therefore, the FIM transforms into the identity matrix under the $\eta$-parameterization:
$$
I_\eta(\eta)=\diag(\sqrt{\theta_1},\ldots,\sqrt{\theta_d})\times \diag\left(\frac{1}{\theta_1},\ldots,\frac{1}{\theta_d}\right)\times\diag(\sqrt{\theta_1},\ldots,\sqrt{\theta_d})=I,
$$
where $I$ denotes the identity matrix.

\begin{figure}
\centering
\includegraphics[width=0.9\textwidth]{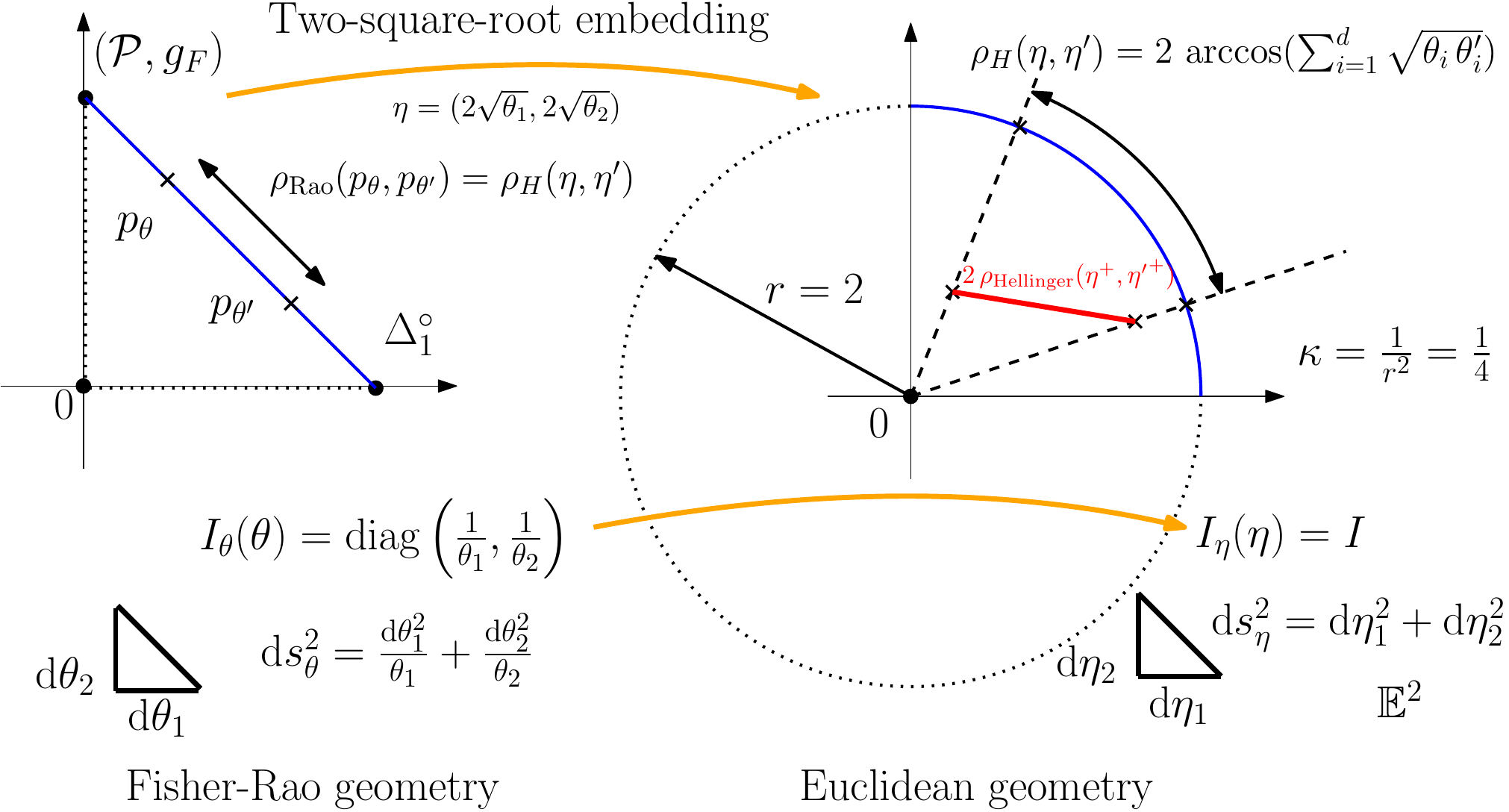}

\caption{Two-square-root embedding of the Bernoulli family onto the positive orthant of the sphere of radius $r=2$.}\label{fig:embedding}
\end{figure}

The transformed parameter space $H=\{\eta(\theta)\st \theta\in\Theta\}$ is the positive orthant of the sphere of radius $r=2$ 
since $\|\eta\|_2=2\sum_{i=1}^d \theta_i=2$ (see Figure~\ref{fig:embedding}).
That is, we have performed an isometric embedding of the Fisher-Rao manifold of the categorical distributions with $d$ atoms into the Euclidean space of dimension $d+1$.
Therefore the Rao distance on the Fisher-Rao manifold of categorical distributions can be computed as the geodesic distance on $H$.
The geodesic distance between $\eta=\eta(p_{\theta})$ and $\eta'=\eta(p_{\theta'})$ on $H$ is
$$
\rho_{H}(\eta,\eta')=2 \angle_{0\eta\eta'}, 
$$
with
$$
\angle_{0\eta\eta'} = \arccos \frac{\eta\cdot\eta'}{\|\eta\|_2 \, \|\eta'\|_2}=
\arccos \left( \frac{4 \sum_{i=1}^d \eta_i\eta_i'}{\|\eta\|_2 \|\eta'\|_2 } \right)
=\arccos \left( \sum_{i=1}^d \\eta_i\, \eta_i'\right).
$$

It follows that the Rao distance between two categorical distributions is
$$
\rho_\calP(p_{\theta},p_{\theta'})=2\,\arccos \left( \sum_{i=1}^d\sqrt{\theta}\sqrt{\theta'}\right).
$$
The term $\sum_{i=1}^d\sqrt{\theta}\sqrt{\theta'}$ is called the Bhattacharyya coefficient.

It follows from the curvature $\kappa=\frac{1}{r^2}$ of a sphere of radius $r$ in $\bbR^d$ that the categorical Fisher-Rao manifold (non-embedded manifold)
has curvature $\kappa=\frac{1}{r^2}=\frac{1}{4}$.

Now, relax the constraint of normalized probabilities $p_\theta$ and consider positive measures $p_\theta^+$ (with $\Theta^+=\bbR_{++}^d$) while keeping the two-square-root embedding.
The extended Rao distance to $\bbR_{++}^d$ becomes:
$$
\rho(p_{\theta_1}^+,p_{\theta_2}^+)=\sqrt{\sum_{i=1}^d (2\sqrt{\theta_1^i}-2\sqrt{\theta_2^i})^2}
=2 \,\sqrt{\sum_{i=1}^d (\sqrt{\theta_1^i}-\sqrt{\theta_2^i})^2 }=2\, \rho_{\Hellinger}(\theta_1,\theta_2),
$$
where
$$
\rho_{\Hellinger}(\theta_1^i,\theta_2^i) = \sqrt{\sum_{i=1}^d (\sqrt{\theta_1^i}-\sqrt{\theta_2^i})^2 }= \|\sqrt{\theta_1}-\sqrt{\theta_2}\|_2,
$$
denotes the Hellinger distance, a metric distance.
The squared Hellinger distance is called the Hellinger divergence and belongs to the class of $f$-divergences 
for the generator $f_\Hellinger(u)=(\sqrt{u}-1)^2$.

Furthermore, the following inequality follows from the embedding of the normalized probabilities on the positive-orthant of the sphere that
$$
\rho(p_{\theta_1},p_{\theta_2})\geq 2\, \rho_{\Hellinger}(\theta_1,\theta_2),
$$
with equality if and only if $\theta_1=\theta_2$.

\subsection{Regular cone manifolds}\label{sec:cone}

\begin{figure}
\centering
\includegraphics[width=0.9\textwidth]{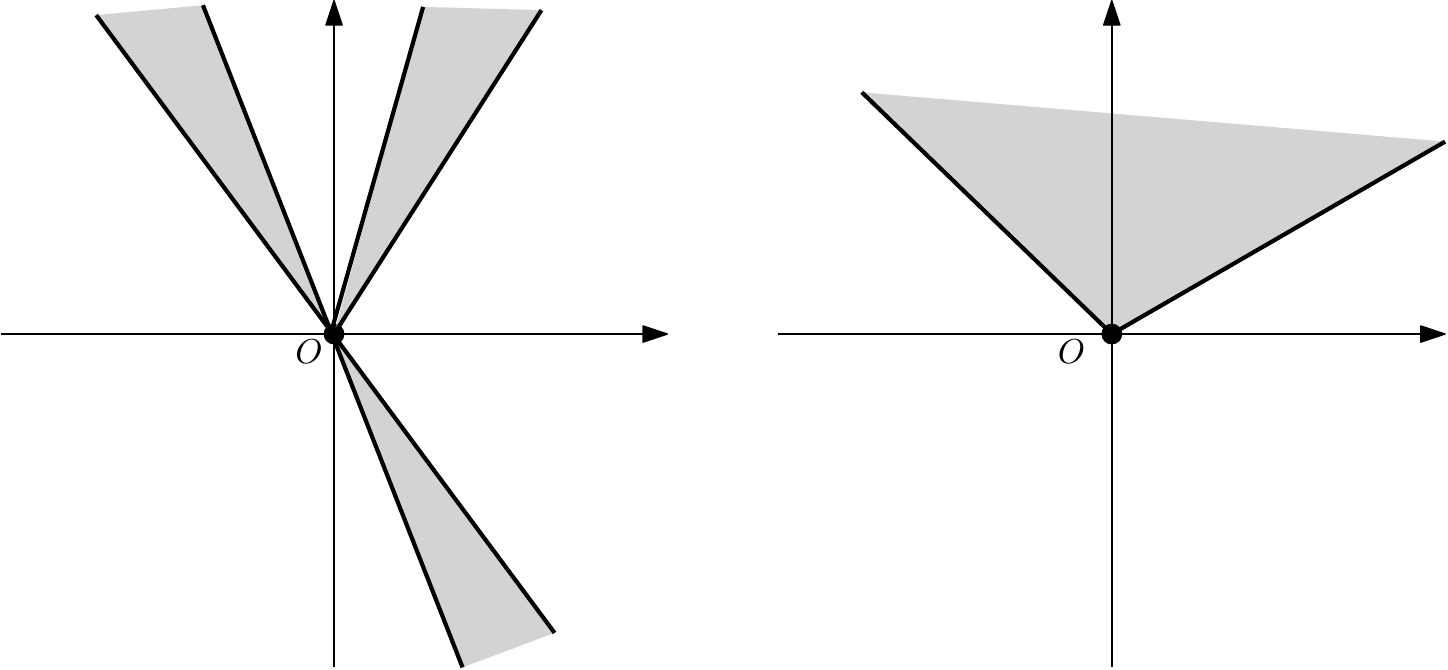}

\caption{Two examples of cones of $\bbR^2$: a non-regular cone (left, non-convex and not pointed) and a regular cone (right).}\label{fig:cone}
\end{figure}

A cone $K\subset\bbR^D$ is a subset such that 
$$
\forall \lambda\geq 0, x\in K, \lambda x\in K.
$$
A cone is said pointed if $K\cap (-K)=0$.
We consider regular cones which are (i) convex and (ii) pointed (i.e., contains no line).
Figure~\ref{fig:cone} displays an example of a non-regular cone (left: non-convex and not pointed) and an example of a regular cone (right)

For a cone $K$, we can associate a dual regular cone $K^*$ defined by
$$
K^*=\cap_{x\in K} \{y\in\bbR^D\st x^\top y\geq 0\}.
$$
When $K$ is regular, we have $(K^*)^*=K$. A cone is self-dual when $K^*=K$.
We can associate to a cone $K$ a characteristic function $\chi_K(x)$ such that for any $x\in K$, we have
$$
\chi_K(x)=\int_{K^*} \exp(-x^\top y) \dy.
$$
Observe the similarity with the partition function of a natural exponential family when the cone is self-dual.
It can be shown that the characteristic function $\chi_K$ is strictly logarithmically convex.
Let $\Aut(K)$ denote the automorphism group of $K$, i.e., the subgroup of the general linear group $\GL(\bbR,d)$ such that
$A\in\Aut(K) \Leftrightarrow A(K)=K$, with $A(K)=\{Ax\ :\ x\in K\}$. The automorphism group can be shown to a be a Lie group~\cite{BarrierConeFormulaCharacteristic-1996}.
A regular cone is said homogeneous if its automorphism group is transitive:
That is, for all $x,y\in K$, there exits $A\in\Aut(K)$ such that $Ax=y$.

It can be shown that 
\begin{equation}\label{eq:chiA}
\chi_K(Ax)=\frac{\chi_K(x)}{|\det(A)|},
\end{equation}
for any $A\in\Aut(K)$.

Since $\chi_K$ is strictly logarithmically convex, let us consider the function
$$
F_K(x)=\log \chi_K(x),
$$
which is strictly convex. Moreover, the function is analytic for homogeneous cones.
We can therefore associate a dually flat space structure to cones~\cite{Shima-2007} (Chapter~4) using $(K,F_K)$.
The induced Riemannian metric $\nabla^2 F(x)$ is invariant under the group automorphism.

Consider a prescribed point $e\in K^\circ$, the interior of $K$.
For any $x\in K^\circ$, let $A_x\in\Aut(K)$ such that $A_xe=x$.
Then using Eq.~\ref{eq:chiA}, we have
$$
F_K(x)=\log \chi_K(e)-\log(|\det(A_x)|).
$$
Since the Bregman generators are defined up to an affine term, we have
$$
F_K(x)\equiv -\log(|\det(A_x)|).
$$

Furthermore, for a homogeneous regular cone, we have~\cite{BarrierConeFormulaCharacteristic-1996} (Theorem 4.4):
$$
F_K(x)\equiv \frac{1}{2}\log \det(\nabla^2 F_K(x)).
$$

For example, consider the non-negative orthant cone $K=\bbR^d_{++}$.
Then we have $F_K(x)=-\sum_{i=1}^d \log x_i$ and $\nabla^2 F_K(x)=\diag\left(\frac{1}{x_1^2},\ldots,\frac{1}{x_d^2}\right)$,
and $\det(\nabla^2 F_K)=\prod_{i=1}^n \frac{1}{x_i^2}$ so that $\frac{1}{2}\log \det(\nabla^2 F_K(x))=-\sum_{i=1}^d \log x_i$.

Consider the cone of symmetric positive-definite matrices of dimension $d\times d$ (SPD cone).
It is a self-dual cone, and the logarithm of the characteristic function is
$F_\SPD(P)=-\frac{d+1}{2}\log \det(P)$.
Notice that the cumulant function of zero-centered multivariate normal distributions $\calN(0,\Sigma)$ is
$$
F_\calN(\Sigma)=-\frac{1}{2}\log \det(\Sigma).
$$
Thus functions $F_\SPD$ and $F_\calN$ differ by a multiplicative factor $d+1$.

G\"uler~\cite{BarrierConeFormulaCharacteristic-1996} investigated a generic way to build universal barrier functions~\cite{lee2021universal} on cones for interior point methods: He proved that $F_K(x)$ for homogeneous cones are self-concordant barrier function (Theorem 4.3~of~\cite{BarrierConeFormulaCharacteristic-1996}).
 We recommend the monograph of Faraut and Kor\'anyi~\cite{faraut1994analysis} for analysis on symmetric cones which are
open convex self-dual homogeneous cones in Euclidean space.

\subsection{Mixture family manifolds}\label{sec:MF}

\subsubsection{Definition}

A mixture family~\cite{IG-2016} $\mathcal{M}$ of order $D$ is defined by $D+1$ linearly independent functions $p_0(x),p_1(x),p_D(x)$ as
$$
\mathcal{M}=\left\{m_\theta(x)=(1-\sum_{i=1}^D \theta_i)p_0(x)+\sum_{i=1}^D p_i(x)\ :\ \theta\in\Delta^\circ_{D} \right\},
$$
where $\Delta^\circ_{D}$ denotes the $D$-dimensional open standard simplex.

Consider a probability space $(\calX,\Sigma_\calX,\mu)$ where $\calX$ denotes the sample space, $\Sigma_\calX$ a $\sigma$-algebra and $\mu$ a positive measure. The set of statistical mixtures with prescribed $D+1$ linear independent components form a mixture family.
Furthermore, it can be shown that the Shannon negentropy
$$
F(\theta)=\int_\calX m_\theta(x)\log m_\theta(x)\dmu(x)
$$
is a strictly convex and smooth function~\cite{MCIG-2019}: A Bregman generator.
Next, we describe in \S\ref{sec:catmix} the discrete mixture family of categorical distributions and point out the difficulty to get the negentropy in closed form in general.
The following section~\ref{sec:CauchyMix} will report an interesting example of analytic continuous mixture of order $1$.

\subsubsection{The categorical distributions: A discrete mixture family}\label{sec:catmix}
Consider the family of categorical distributions as a mixture family
$$
\mathcal{M}=\left\{m_\theta(x)=(1-\sum_{i=1}^D \theta_i)\delta_{x_0}(x)+\sum_{i=1}^D \delta_{x_D}(x)\ :\ \theta\in\Delta^\circ_{D} \right\},
$$
where $\delta_{x_i}(x)=\delta(x-x_i)=1$ iff $x=x_i$ and $0$ when $x\not=x_i$. 
The functions $\delta_{x_i}$ are called Dirac distributions and are linearly independent provided that $x_i\not=x_j$ for any $i\not=j$.
The Shannon negentropy is
$$
F(\theta)=\sum_{x\in\{x_0,x_1,\ldots, x_D\}} m_\theta(x)\log m_\theta(x)=\sum_{i=0}^D \theta_i\log\theta_i.
$$

The discrete mixture family of categorical distributions can be extended to continuous mixtures with mixture components having pairwise disjoint support $\calX_i\cap\calX_j=\emptyset$ for all $i\not=j$.
We have
\begin{eqnarray*}
F(\theta) &=&
\int_{\calX}m_\theta(x)\log m_\theta(x)\dmu(x),\\
&=&\sum_{i=1}^D \int_{\calX_i} m_\theta(x)\log m_\theta(x)\dmu(x)
+\left(1-\sum_{i=1}^D \theta_i\right) \int_{\calX_0} m_\theta(x)\log m_\theta(x)\dmu(x),\\
&=& \sum_{i=1}^D \int_{\calX_i} (\theta_i p_i(x))\log (\theta_i p_i(x))\dmu(x) + \left(1-\sum_{i=1}^D \theta_i\right)  
\int_{\calX_0} p_0(x)\log p_0(x)\dmu(x),\\
&=& \sum_{i=1}^D \theta_i(\log\theta_i + I_i) + \left(1-\sum_{i=1}^D \theta_i\right)   I_0,
\end{eqnarray*}
where $I_j=\int_{\calX_j} p_j(x)\log p_j(x)\dmu(x)$ is Shannon negentropy of component $p_j$.
Thus we have
$$
F(\theta) = \sum_{i=1}^D \theta_i\log\theta_i +  \sum_{i=1}^D \theta_i (I_i-I_0). 
$$
Since Bregman generators are equivalent affine terms, it follows that
$$
F(\theta) \equiv \sum_{i=1}^D \theta_i\log\theta_i.
$$
When $p_j(x)=\delta_{x_j}(x)$ with $\calX_j=\{x_j\}$, we recover the discrete mixture family of categorical distributions.

\section{Information geometry of the mixture family of two distinct Cauchy distributions}\label{sec:CauchyMix}

\subsection{Cauchy mixture family of order $1$}

The probability density function of a Cauchy distribution with location parameter $l$ and scale parameter $s>0$ is
$$
p_{l,s}(x) := \frac{1}{\pi s\left(1+\left(\frac{x-l}{s}\right)^2\right)}=\frac{s}{\pi (s^2+(x-l)^2)}.
$$
The family of Cauchy distributions form a location-scale family 
$$
\calC=\left\{p_{l,s}(x):=\frac{1}{s} p\left(\frac{x-l}{s}\right) \st (l,s)\in \bbR\times\bbR_{++}\right\},
$$ 
with standard Cauchy distribution 
$$
p_{0,1}(x) := \frac{1}{\pi(1+x^2)}.
$$

Consider the mixture family~\cite{IG-2016,EIG-2020} induced by two distinct Cauchy distributions $p_{l_0,s_0}$ and $p_{l_1,s_1}$:
\begin{eqnarray*}
\calM:= \left\{m_\theta(x) := (1-\theta)p_{l_0,s_0}(x)+\theta p_{l_1,s_1}(x)\ :\ \theta\in(0,1)\right\}.
\end{eqnarray*}

\begin{figure}
\centering


\begin{tabular}{lll}
\includegraphics[width=0.3\textwidth]{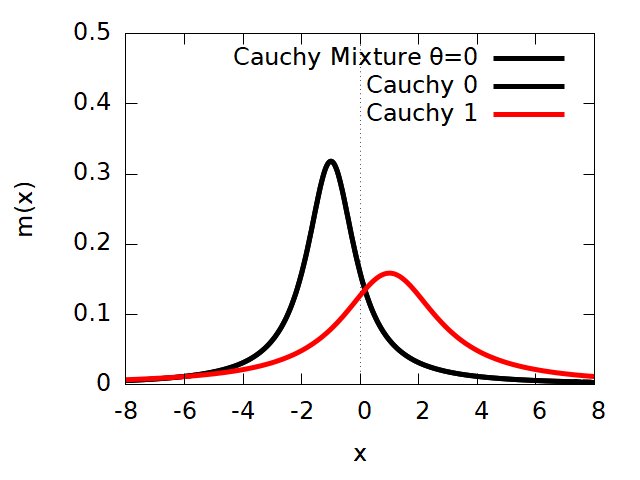} &
\includegraphics[width=0.3\textwidth]{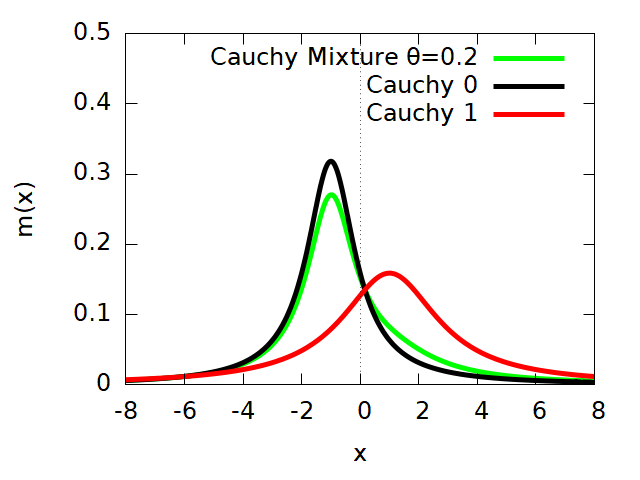} &
\includegraphics[width=0.3\textwidth]{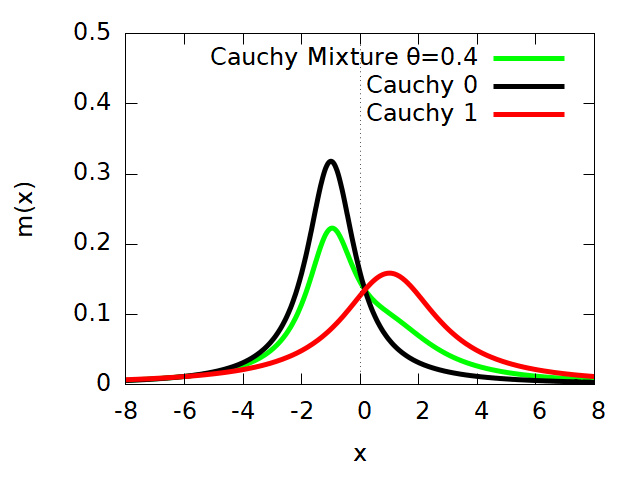} \\
\includegraphics[width=0.3\textwidth]{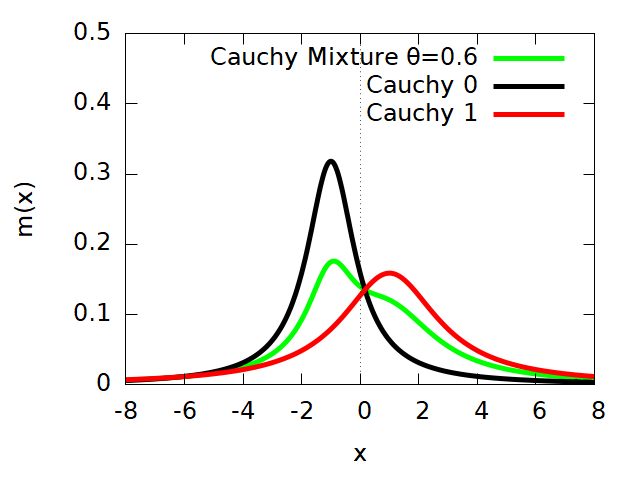} &
\includegraphics[width=0.3\textwidth]{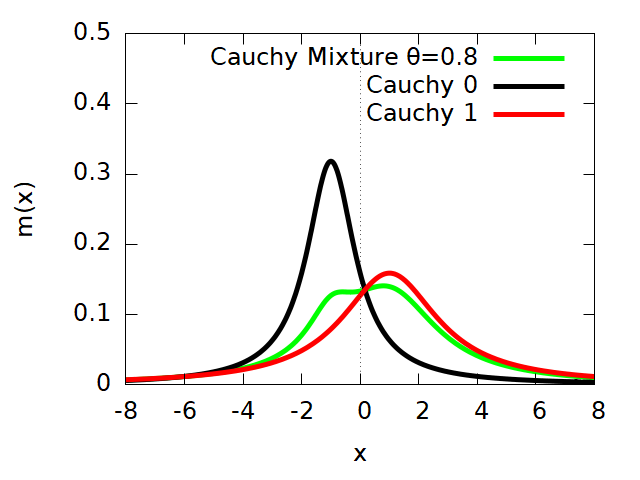} &
\includegraphics[width=0.3\textwidth]{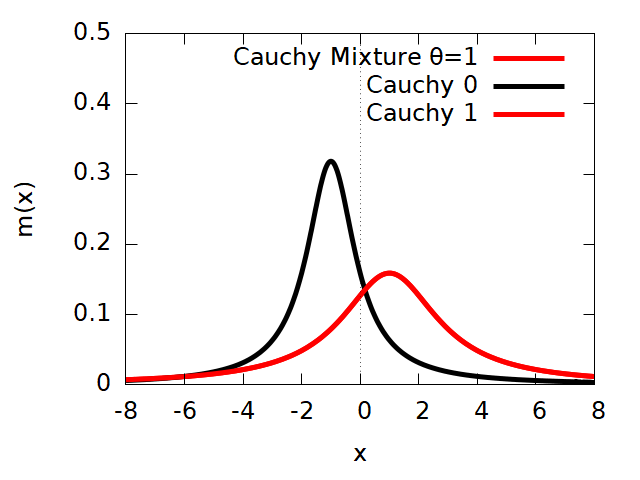} 
\end{tabular}

\caption{An example of a Cauchy mixture $\{m_\theta(x)\ :\ \theta\in (0,1)\}$ of two components with $(l_0,s_0)=(-1,1)$ and $(l_1,s_1)=(1,2)$: From top left to bottom right: $m_\theta(x)$ for $\theta\in(0,0.2,0.4,0.6,0.8,1)$. Notice that $m_0(x)$ and $m_1(x)$ do not belong to the mixture family since $\theta\in (0,1)$.}
\label{fig:excauchymix}
\end{figure}

Figure~\ref{fig:excauchymix} displays an example of a Cauchy mixture of two components with $(l_0,s_0)=(-1,1)$ and $(l_1,s_1)=(1,2)$.

Because the mixture $m_\theta$ is a convex combination of two prescribed Cauchy components, these statistical mixtures have also been called $w$-mixtures in~\cite{wmixtures-2018} (stands for weight mixtures).

The  Kullback-Leibler divergence  between two continuous probability densities $p(x)$ and $q(x)$ is
\begin{eqnarray*}
D_\KL[p:q] &=&\int p(x)\log \frac{p(x)}{q(x)}\mathrm{d}x,\\
&=& h^\times[p:q]-h[p],
\end{eqnarray*}
where $h^\times[p:q]=-\int p(x)\log q(x)\mathrm{d}x$ is the cross-entropy, and $h[p]=h^\times[p:p]=-\int p(x)\log p(x)\mathrm{d}x$ is the differential entropy.

In~\cite{fdivCauchy-2021}, the following closed-form formula (using complex analysis in $\bbC$) was proven for the Kullback-Leibler divergence between a Cauchy density and a mixture of two Cauchy densities:
\begin{eqnarray}
\lefteqn{D_\KL[p_{l_0,s_0}:m_\theta]=} \label{eq:klpmix} \\ 
&&{  \log\left(\frac{(l_0 - l_1)^2 + (s_0+s_1)^2}{(1-\theta)(s_0^2 + s_1^2 +(l_0 - l_1)^2)+ 2\theta s_0 s_1 +
2\sqrt{s_0^2 s_1^2 + s_0 s_1 ((s_0-s_1)^2+(l_0 - l_1)^2)\theta (1-\theta)}}\right)}.  \nonumber
\end{eqnarray}

Define the skewed $\theta$-Jensen-Shannon divergence~\cite{Lin-1991} for $\theta\in(0,1)$:
\begin{eqnarray}
D_{\JS,\theta}(p_{l_0,s_0}:p_{l_1,s_1}) &:=& (1-\theta)D_\KL(p_{l_0,s_0}:m_\theta)+\theta D_\KL(p_{l_1,s_1}:m_\theta),\\
  &=& h[(1-\theta)p_{l_0,s_0}+\theta p_{l_1,s_1}]-((1-\theta)h[p_{l_0,s_0}]+\theta h[p_{l_1,s_1}]).
\end{eqnarray}

Since $D_\KL(p_{l_1,s_1}:m_\theta)=D_\KL(p_{l_1,s_1}:m'_{1-\theta})$ with
$m'_\theta(x) := (1-\theta)p_{l_1,s_1}(x)+\theta p_{l_0,s_0}(x)$, we can use Eq.~\ref{eq:klpmix} to calculate $D_\KL(p_{l_1,s_1}:m_\theta)$.
Using the fact that the Shannon entropy for the Cauchy density $p_{l,s}$ is $h[p_{l,s}]=\log(4\pi s)$~\cite{KLCauchy-2019},
 we thus get the Shannon differential entropy of a $2$-component Cauchy mixture in closed form:

\begin{prop}[Entropy of a $2$-Cauchy mixture]
The differential entropy of a mixture of two Cauchy distributions is available in closed-form:
\begin{equation}
h[m_\theta]=h[(1-\theta)p_{l_0,s_0}+\theta p_{l_1,s_1}] = D_{\JS,\theta}(p_{l_0,s_0}:p_{l_1,s_1}) + ((1-\theta)h[p_{l_0,s_0}]+\theta h[p_{l_1,s_1}]).
\end{equation}
\end{prop}

Since the differential entropy of a location-scale density can be expressed as $h[p_{l,s}]=\log s+h[p_{0,1}]$ (by a change of variable in the definite integral of Shannon entropy), we have
\begin{equation}\label{eq:hmtheta}
h[m_\theta] = D_{\JS,\theta}(p_{l_0,s_0}:p_{l_1,s_1})  + \theta\log\frac{s_1}{s_0} + \log(4\pi s_0),
\end{equation}
since $h[p_{0,1}]=\log4\pi$~\cite{KLCauchy-2019}.

Using symbolic computing detailed in Appendix~\ref{sec:maxima}, we get a closed-form formula for Eq.~\ref{eq:hmtheta}.
Without loss of generality\footnote{Indeed, one can consider the action of the location-scale group otherwise.}, we may assume $(l_0,s_0)=(0,1)$ and $(l_1,s_1)=(l,s)$.
We get
\begin{eqnarray}\label{eq:canohmtheta}
h[m_\theta] &=& 
\theta\,\log \left({{\left(s+1\right)^2+l^2}\over{2\,\sqrt{
 \left(\left(s-1\right)^2+l^2\right)\,s\,\left(1-\theta\right)\,
 \theta+s^2}+\left(s^2+l^2+1\right)\,\theta+2\,s\,\left(1-
 \theta\right)}}\right)\nonumber\\
&&+\left(1-\theta\right)\,\log \left({{
 \left(s+1\right)^2+l^2}\over{2\,\sqrt{\left(\left(1-s\right)^2+l^2
 \right)\,s\,\left(1-\theta\right)\,\theta+s^2}+2\,s\,\theta
 +\left(s^2+l^2+1\right)\,\left(1-\theta\right)}}\right)\nonumber\\
&&+\theta\log s+\log(4\pi).
\end{eqnarray}

The general formula for the differential entropy of the mixture of two Cauchy distributions is
{\footnotesize
\begin{eqnarray}
h[m_\theta] &=&  \theta\,\log \left({{\left(s_{1}+s_{0}\right)^2+\left(l_{1}-
 l_{0}\right)^2}\over{2\,\sqrt{s_{0}\,s_{1}\,\left(\left(s_{1}-s_{0}
 \right)^2+\left(l_{1}-l_{0}\right)^2\right)\,\left(1-\theta
 \right)\,\theta+s_{0}^2\,s_{1}^2}+\left(s_{1}^2+s_{0}^2+\left(
 l_{1}-l_{0}\right)^2\right)\,\theta+2\,s_{0}\,s_{1}\,\left(1-
 \theta\right)}}\right)\nonumber\\
&&+\left(1-\theta\right)\,\log \left({{
 \left(s_{1}+s_{0}\right)^2+\left(l_{0}-l_{1}\right)^2}\over{2\,
 \sqrt{s_{0}\,\left(\left(s_{0}-s_{1}\right)^2+\left(l_{0}-l_{1}
 \right)^2\right)\,s_{1}\,\left(1-\theta\right)\,\theta+s_{0}^2
 \,s_{1}^2}+2\,s_{0}\,s_{1}\,\theta+\left(s_{1}^2+s_{0}^2+\left(
 l_{0}-l_{1}\right)^2\right)\,\left(1-\theta\right)}}\right)\nonumber\\
&&+ \theta\log\frac{s_1}{s_0} + \log(4\pi s_0). \label{eq:hgenCauchy}
\end{eqnarray}
}

In~\cite{fdivCauchy-2021}, the following closed-form formula was reported for the Jensen-Shannon divergence when $\theta=\frac{1}{2}$:
\begin{eqnarray*}
D_\JS(p_{l_0,s_0}:p_{l_1,s_1}) &=& h[m_{\frac{1}{2}}]-\frac{1}{2}\left(h[p_{l_0,s_0}+h[p_{l_1,s_1}]\right),\\
&=& \log\left( \frac{2\sqrt{(l_0 - l_1)^2 + (s_0+s_1)^2}}{\sqrt{(l_0 - l_1)^2 + (s_0+s_1)^2} + 2\sqrt{s_0s_1}}\right).
\end{eqnarray*}


Consider the negentropy as a univariate Bregman generator:
$$
F(\theta):=-h[m(\theta)].
$$
It can be proven that $F(\theta)$ is a strictly convex function~\cite{MCIG-2019}.

\begin{figure}
\centering

\begin{tabular}{cc}
\includegraphics[width=0.35\textwidth]{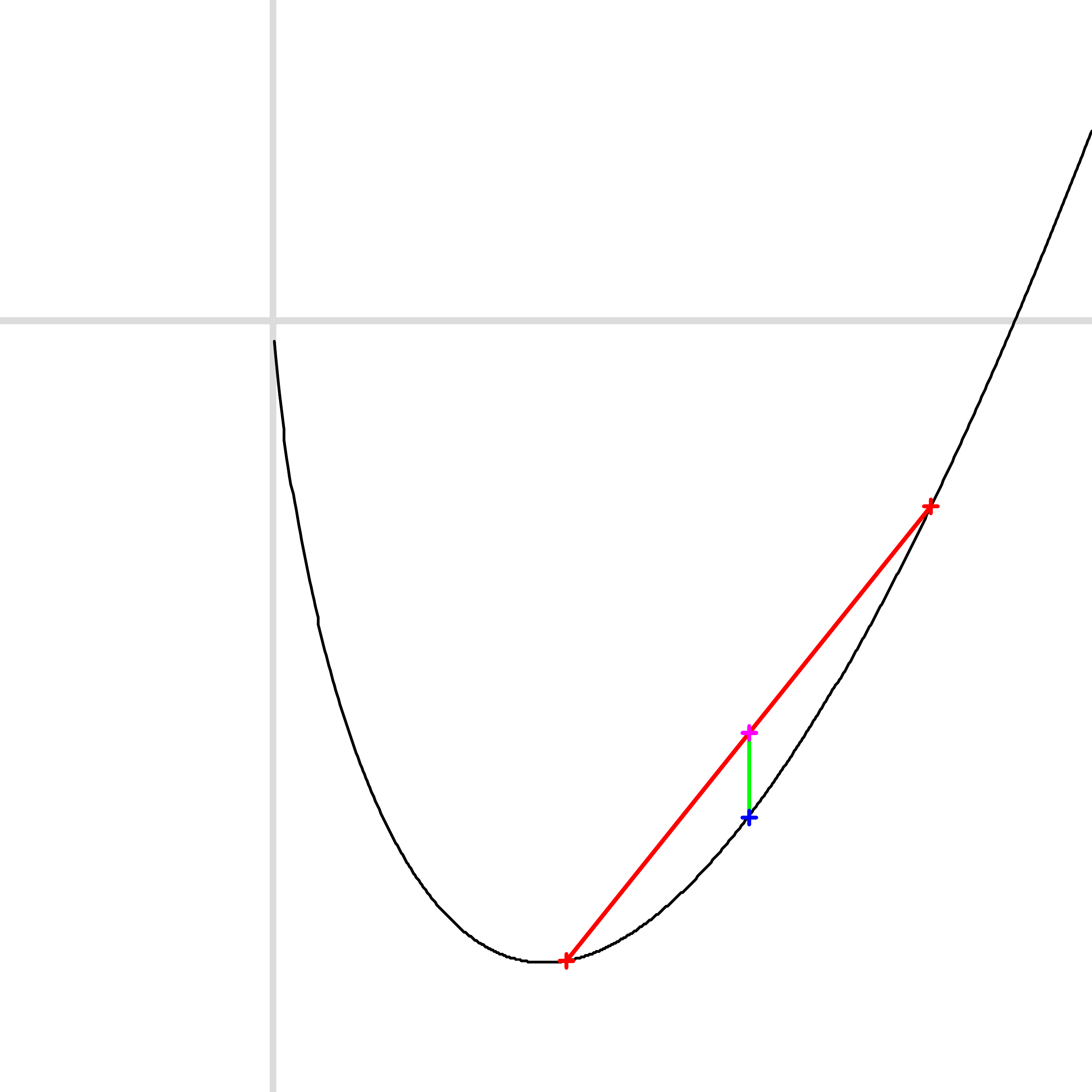} &
\includegraphics[width=0.35\textwidth]{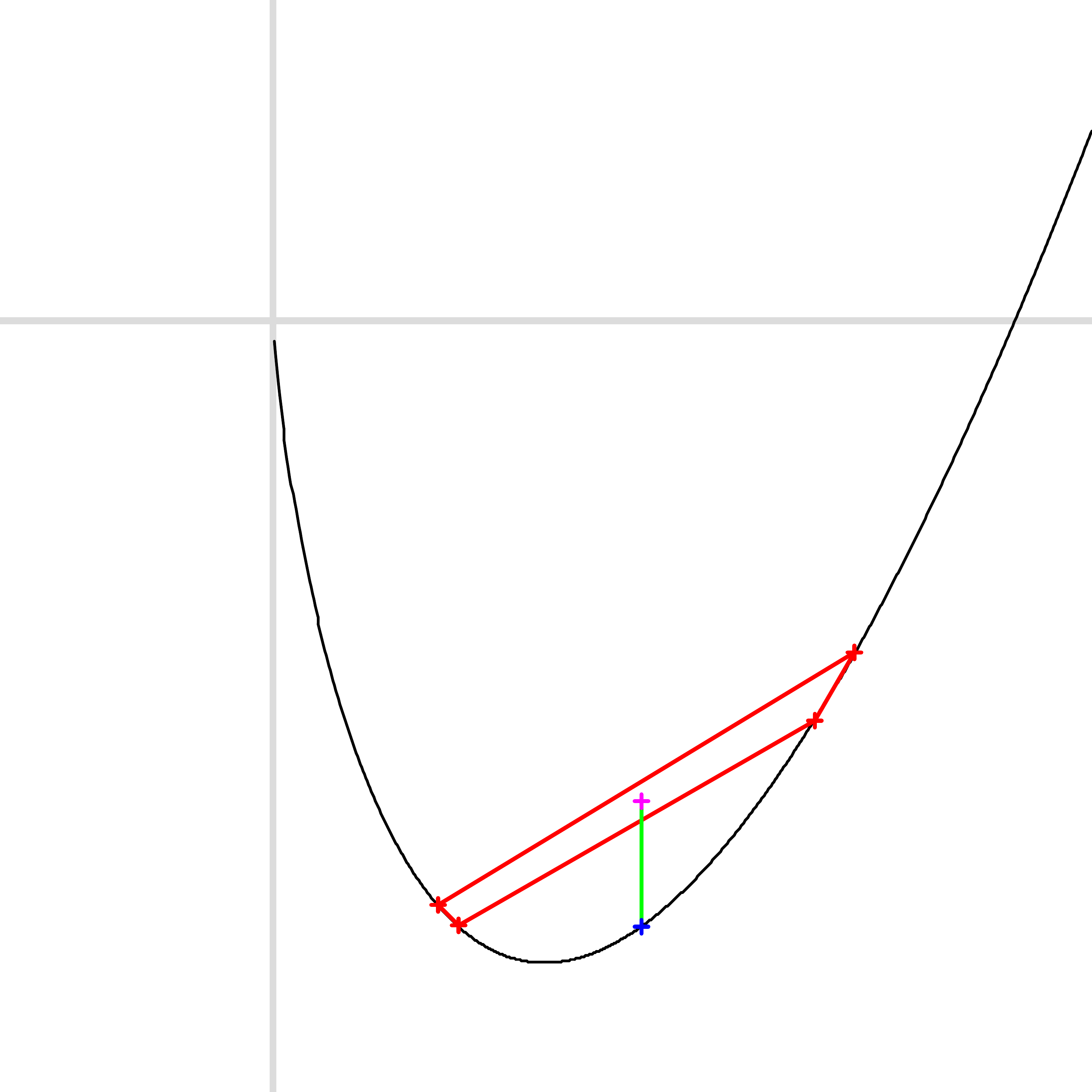} \\
Jensen divergence ($D=2$) & Jensen diversity ($D>2$)
\end{tabular}

\caption{Illustration of the Jensen divergence between two parameters (left, $D=2$) and   its generalization to a set of $D$ parameters called the Jensen diversity (right, $D=4$) as the vertical gap between $E[F(\theta)]-F(E[\theta])\geq 0$ (shown in green).
 }\label{fig:JensenDivergenceDiversity}
\end{figure}

\begin{figure}
\centering
\includegraphics[width=0.85\textwidth]{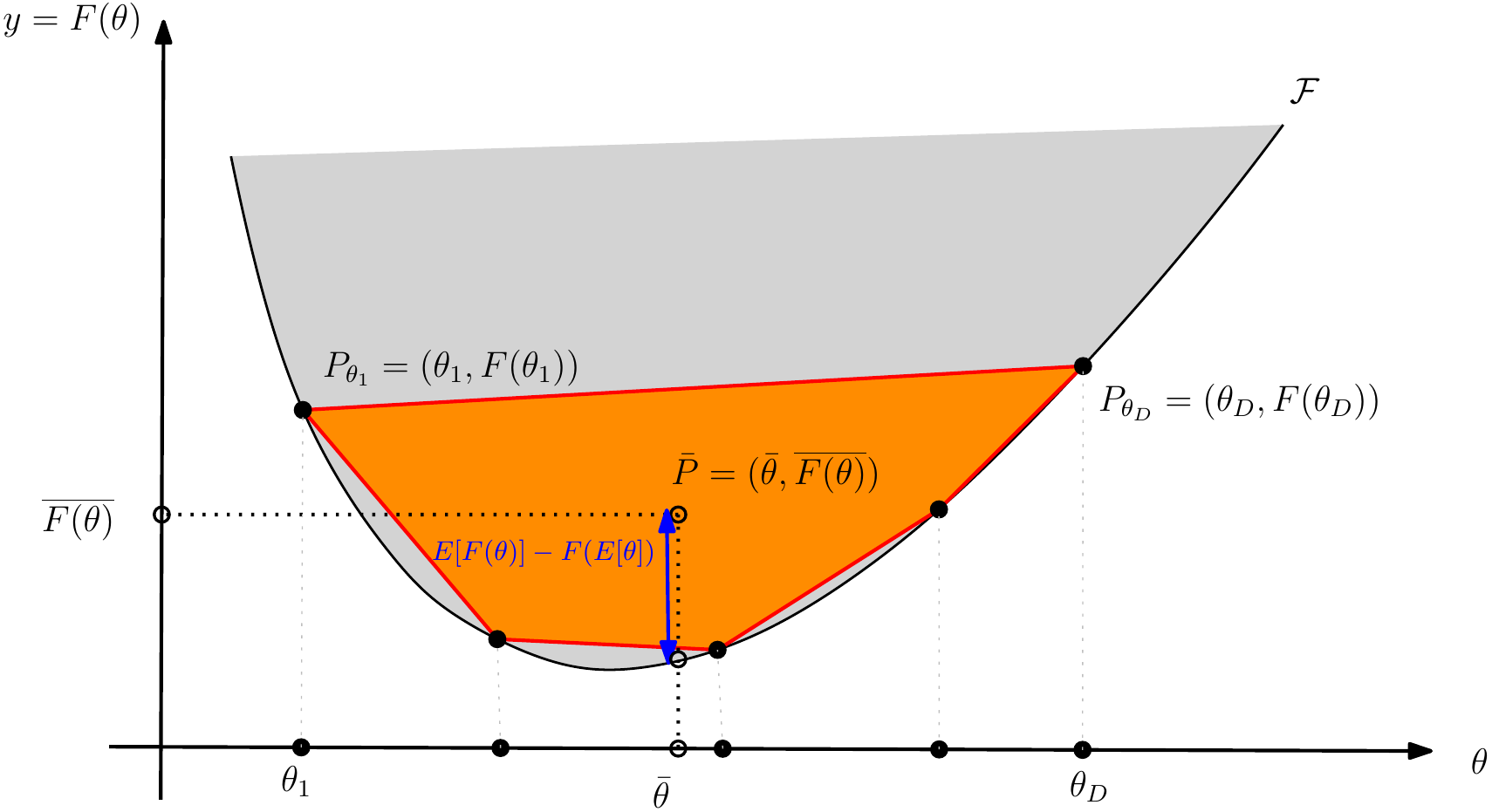}

\caption{Visual proof of Jensen's inequality: The center of mass $(\bar\theta,\frac{1}{D}\sum_i F(\theta_i))$ of the points $\theta_1,\ldots,\theta_D$ is necessarily contained in the convex hull $\mathcal{CH}$ of the points $(\theta_1,F(\theta_1)),\ldots, (\theta_2,F(\theta_2))$.
Since the convex hull $\mathcal{CH}$ is included in the epigraph of function $F(\theta)$, we have $\frac{1}{D}\sum_i F(\theta_i)\geq F\left(\frac{1}{D}\sum_i\theta_i\right)$.
 }\label{fig:JensenDiversityCH}
\end{figure}

The skewed Jensen-Shannon divergence is thus interpreted as a Jensen divergence for the smooth and strictly convex generator $F(\theta):=-h[m(\theta)]$:
\begin{eqnarray*}
D_{\JS,\theta}(p_{l_0,s_0}:p_{l_1,s_1}) &=& h[(1-\theta)p_{l_0,s_0}+\theta p_{l_1,s_1}]-((1-\theta)h[p_{l_0,s_0}]+\theta h[p_{l_1,s_1}]),\\
&=& (1-\theta)F(0)+\theta F(1) - F(\theta),\\
&=& J_{F,\theta}(0:1),
\end{eqnarray*}
where
$$
J_{F,\alpha}(\theta_1:\theta_2):= (1-\alpha)F(\theta_1)+\alpha F(\theta_2)-F((1-\alpha)\theta_1+\alpha\theta_2)
$$
is called the skewed Jensen divergence (or Burbea-Rao divergence~\cite{BR-2011,IG-2016}) induced by generator $F(\theta)$.
The Jensen divergence between two parameters can be extended to the Jensen diversity between a set of $n\geq 2$ weighted parameters $\theta_1,\ldots, \theta_n$ with corresponding weights $w_1,\ldots, w_n$ (such that $w_i>0$ and $\sum_i w_i=1$) as follows:
$$
J_F(\theta_1,\ldots,\theta_n;w_1,\ldots, w_n)=(\sum_{i=1}^n w_i F(\theta_i)) - F\left(\sum_{i=1}^n w_i\theta_i\right).
$$ 
Clearly, we have $J_{F,\alpha}(\theta_1:\theta_2)=J_F(\theta_1,\theta_2;1-\alpha,\alpha)$. See Figure~\ref{fig:JensenDivergenceDiversity}.
The fact the Jensen diversities are proper divergences (i.e., $J_{F,\alpha}(\theta_1:\theta_2)\geq 0$ with equality iff $\theta_1=\ldots=\theta_n$) stems from Jensen's inequality. A geometric proof of the discrete Jensen's inequality is given in Figure~\ref{fig:JensenDiversityCH}.

Since the Bregman generator is available in closed form, we can also calculate the dual coordinate~\cite{EIG-2020} by differentiation:
$$
\eta(\theta):=F'(\theta).
$$

It was shown in~\cite{MCIG-2019} that the dual convex conjugate $F^*(\eta)$ amounts to calculate the cross-entropy between $p_0$ and the mixture $m_\theta$ (which can also be parameterized equivalently as $m_\eta$):
$$
F^*(\eta)=-\int p_0(x)\log m_\eta(x) \dmu(x).
$$

Using the formula of Eq.~\ref{eq:klpmix} and the fact that
$$
D_\KL(p_{l_0,s_0}:m_\theta) = h^\times[p_{l_0,s_0}:m_\theta]-h[p_{l_0,s_0}],
$$
we deduce a closed-form formula for the cross-entropy  between $p_0$ and the mixture $m_\theta$ since $h[p_{l_0,s_0}]=\log(4\pi s_0) $.

\begin{prop}[Cross-entropy closed-form formula]
The cross-entropy  between $p_0$ and the mixture $m_\theta$ is:
\begin{equation}
h^\times[p_{l_0,s_0}:m_\theta]=D_\KL(p_{l_0,s_0}:m_\theta) + \log(4\pi s_0). 
\end{equation}
\end{prop}

Thus the dual convex conjugate $F^*$ can be expressed in closed-form using the $\theta$-coordinate system: 
$F^*(\eta(\theta))=h^\times[p_{l_0,s_0}:m_\theta]$.

It was shown in~\cite{EIG-2020} how to reconstruct the statistical divergence corresponding to the Bregman divergence $B_F$ induced by the  Bregman generator $F$. When the Bregman generator is the negentropy of a mixture family, the Kullback-Leibler divergence is reconstructed~\cite{EIG-2020}:

Thus we have
\begin{eqnarray}
D_\KL[m_{\theta_1}:m_{\theta_2}] &=& B_F(\theta_1:\theta_2),\\
&=& F(\theta_1)-F(\theta_2)-(\theta_1-\theta_2) F'(\theta_2),\\
&=& F(\theta_1)+F^*(F'(\theta_2))-\theta_1 F'(\theta_2).
\end{eqnarray}

This last formula is thus available in closed-form since $F(\theta)$ is available in closed form.
We may also derive this formula as follows:
\begin{eqnarray}
D_\KL[m_{\theta_1}:m_{\theta_2}] &=& h^\times[m_{\theta_1}:m_{\theta_2}]-h[m_1]\\
&=& (1-\theta_1)h^\times[p_0:m_{\theta_2}] + \theta_1 h^\times[p_1:m_{\theta_2}]-h[m_1],
\end{eqnarray}
and use the above closed-form formula.

We show in the Appendix how to report an explicit closed-form formula with respect to $l_0, s_0, l_1, s_1, \theta_1$ and $\theta_2$.

It follows that the Jeffreys divergence:
\begin{eqnarray}
D_J[m_{\theta_1}:m_{\theta_2}] &:=& D_\KL[m_{\theta_1}:m_{\theta_2}]+D_\KL[m_{\theta_2}:m_{\theta_1}],\\
&=& (\theta_2-\theta_1)(\eta_2-\eta_1),
\end{eqnarray}
 and the Jensen-Shannon divergence: 
\begin{eqnarray}
D_\JS[m_{\theta_1}:m_{\theta_2}] &:=&  \frac{1}{2}\left(D_\KL\left[m_{\theta_1}:\frac{m_{\theta_1}+m_{\theta_2}}{2}\right] +
D_\KL\left[m_{\theta_2}:\frac{m_{\theta_1}+m_{\theta_2}}{2}\right]\right),\\
&=& \frac{1}{2}\left(D_\KL\left[m_{\theta_1}:m_{\frac{\theta_1+\theta_2}{2}}\right] + D_\KL\left[m_{\theta_2}:m_{\frac{\theta_1+\theta_2}{2}}\right]\right),\\
&=& h\left[m_{\frac{\theta_1+\theta_2}{2}}\right]-\frac{1}{2}\left(h[m_{\theta_1}]+h[m_{\theta_2}]\right).
\end{eqnarray}
are also available in closed-form~\cite{wmixtures-2018} (using Eq.~\ref{eq:hgenCauchy}).

\begin{prop}
The Kullback-Leibler divergence, Jeffreys divergence and Jensen-Shannon divergence 
between any two mixtures of Cauchy distributions with prescribed two distinct components
can be calculated in closed forms.
\end{prop}

Notice that closed-form formula for the Kullback-Leibler divergence
 and the Jensen-Shannon divergence between Cauchy distributions were reported in~\cite{fdivCauchy-2021}.

To express $F^*(\eta)$ using the $\eta$-coordinate, we need first to calculate $\theta(\eta)={F^*}'(\eta)=(F')^{-1}(\eta)$, i.e.,
 inverse $F'(\theta)$, and then apply the Legendre formula:
$$
F^*(\eta)=\theta(\eta)\eta-F(\theta(\eta)).
$$ 

By fixing $(l_0,s_0)$ and $(l_1,s_1)$, we can obtain $\theta(\eta)$ in closed form (and therefore $F^*(\eta)$).
We report an example in the next section.

\subsection{An analytic example}

Let us illustrate the construction of dually flat space for the family of mixtures of two Cauchy distributions with prescribed parameters
  $(l_0,s_0)=(0,1)$ and $(l_1,s_1)=(1,1)$.
We have the following induced generator:
\begin{eqnarray}
F_{0,1,1,1}(\theta)&=&-h[(1-\theta)p_{0,1}+\theta p_{1,1}],\nonumber\\
&=&\theta\log\left(\frac{2\sqrt{1+\theta-\theta^2}+\theta+2}{2\sqrt{1+\theta-\theta^2}-\theta+3}\right)
+\log\left(\frac{2\sqrt{1+\theta-\theta^2}-\theta+3}{20\pi}\right). \label{eq:exF}
\end{eqnarray}

\begin{figure}%
\centering
\includegraphics[width=0.5\columnwidth]{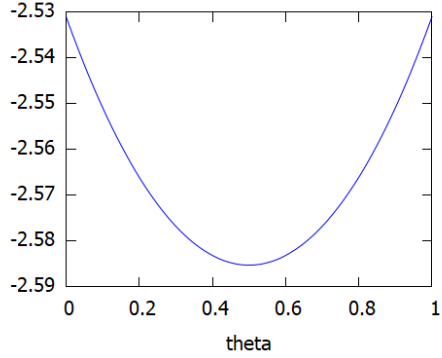}%
\caption{Plot of the strictly convex and smooth Bregman generator $F_{0,1,1,1}(\theta)$.}%
\label{fig:plotF}%
\end{figure}

Figure~\ref{fig:plotF} plots the potential function $F(\theta)$ for $\theta\in(0,1)$.

We can therefore calculate in closed-form the derivative of $F_{0,1,1,1}(\theta$):
\begin{equation}
\eta(\theta)=F_{0,1,1,1}'(\theta)=\log \left( \frac{2\sqrt{1+\theta-\theta^2}+\theta+2}{2\sqrt{1+\theta-\theta^2}-\theta+3} \right).
\label{eq:exgradF}
\end{equation}

The Bregman divergence $B_{F_{0,1,1,1}}(\theta_1:\theta_2)$ is

\begin{eqnarray}
&&B_{F_{0,1,1,1}}(\theta_1:\theta_2) = D_\KL[m_{\theta_1}:m_{\theta_2}],\nonumber\\
&=& \theta_1\log\frac{(2\sqrt{1+\theta_1-\theta_1^2}+\theta_1+2)(2\sqrt{1+\theta_2-\theta_2^2}-\theta_2+3)}{
(2\sqrt{1+\theta_1-\theta_1^2}-\theta_1+3)(2\sqrt{1+\theta_2-\theta_2^2}+\theta_2+2)}
+\log\frac{2\sqrt{1+\theta_1-\theta_1^2}-\theta_1+3}{2\sqrt{1+\theta_2-\theta_2^2}-\theta_2+2}.\label{eq:exBD}
\end{eqnarray}

 
Now, let us calculate the inverse function $\theta(\eta)$ in closed-form using a computer algebra software:
\begin{equation}
\theta(\eta)={F_{0,1,1,1}^*}'(\eta)= \frac{5 \exp(2\eta)+ 2\sqrt{5}\sqrt{\exp(3\eta)-2\exp(2\eta)+\exp(\eta)}-3\exp(\eta)}{5\exp(2\eta)-6\exp(\eta)+5}.\label{eq:exgradG}
\end{equation}

It follows that the dual potential function $F^*(\eta)$ is available in closed form (see Appendix for the closed-form formula):
\begin{eqnarray*}
F^*_{0,1,1,1}(\eta) &=&\theta(\eta)\eta-F_{0,1,1,1}(\theta(\eta)).
\end{eqnarray*}

We have the dual potential expressed using the $\theta$-coordinate as:
$$
F^*_{0,1,1,1}(\eta(\theta))=\log \left(\frac{20\,\pi}{2\sqrt{1+\theta-\theta^2}-\theta+3}\right).
$$

In Appendix, we show how to report in closed-form $F^*(\eta)$ by expanding the formula: This yields a long formula which we do not paste here, from which we can also obtain ${F^*_{0,1,1,1}}'(\eta)$.

The metric tensor $\leftsup{F_{0,1,1,1}}g$ can be calculated in closed form in the $\theta$-coordinate system using the second derivative:
$[\leftsup{F_{0,1,1,1}}g]_\theta=F_{0,1,1,1}''(\theta)$.
We get (see Appendix):
\vskip 0.3cm
\noindent
\scalebox{.75}{
$[\leftsup{F_{0,1,1,1}}g]_\theta=-{{-\theta^8+4\,\theta^7+\sqrt{-\theta^2+\theta+1}\,
 \left(7\,\theta^6-21\,\theta^5-35\,\theta^4+105\,\theta^
 3+56\,\theta^2-112\,\theta-64\right)+19\,\theta^6-71\,
 \theta^5-30\,\theta^4+183\,\theta^3+40\,\theta^2-144\,
 \theta-64}\over{-\theta^{10}+5\,\theta^9+\sqrt{-\theta^2
 +\theta+1}\,\left(8\,\theta^8-32\,\theta^7-40\,\theta^6+
 232\,\theta^5+16\,\theta^4-456\,\theta^3-48\,\theta^2+
 320\,\theta+128\right)+23\,\theta^8-122\,\theta^7+\theta
 ^6+445\,\theta^5-127\,\theta^4-640\,\theta^3+32\,\theta^
 2+384\,\theta+128}}. 
$
}
\vskip 0.15cm

Thus we reported in closed-form all necessary equations to implement the mixture family of two distinct Cauchy distributions (Eq.~\ref{eq:exF}, Eq.~\ref{eq:exgradF}, Eq.~\ref{eq:exBD}, Eq.~\ref{eq:exgradG}).

The Cauchy distributions are Student's $t$-distributions for $\nu=1$ degree of freedom.
When the degrees of freedom tend to infinity, the Cauchy distributions tend to normal distributions.

\vskip 0.5cm

\noindent {\bf Acknowledgments.} The author is very indebted to Professor Kazuki Okamura (Shizuoka University, Japan) which collaborated with the author on the study of $f$-divergences between Cauchy distributions~\cite{fdivCauchy-2021}. 
In this work, all the $\alpha$-geometries~\cite{IG-2016} of the family of Cauchy distributions are proven to coincide with the Fisher-Rao hyperbolic geometry of the Cauchy family due to the symmetric property of the $f$-divergences for the Cauchy family.

\appendix

\section{Symbolic computing notebook in {\sc Maxima}}\label{sec:maxima}
The following notebook can be executed using the computer algebra system {\sc Maxima}\footnote{\url{https://maxima.sourceforge.io/}} for symbolic computations:

First, we initialize the closed-form formula:
\begin{lstlisting}[breaklines,backgroundcolor = \color{lightgray},basicstyle=\small]
KLCauchy(l0,s0,l1,s1,theta):=log(  ( (l0-l1)**2+(s0+s1)**2) / ( (1-theta)*(s0*s0+s1*s1+(l0-l1)**2) + 2*theta*s0*s1+2*sqrt(s0*s0*s1*s1+s0*s1*((s0-s1)**2+(l0-l1)**2)*theta*(1-theta)) ));
JSCauchy(l0,s0,l1,s1,theta):=(1-theta)*KLCauchy(l0,s0,l1,s1,theta) + theta*KLCauchy(l1,s1,l0,s0,1-theta);
hCauchy(l,s):=log(4*%pi*s);
hmixCauchy(l0,s0,l1,s1,theta) :=   JSCauchy(l0,s0,l1,s1,theta)+((1-theta)*hCauchy(l0,s0)+theta*hCauchy(l1,s1))  ;
crossentropyCauchy(l0,s0,l1,s1,theta) := KLCauchy(l0,s0,l1,s1,theta)+hCauchy(l0,s0);
KLmix(l0,s0,l1,s1,theta1,theta2):=(1-theta1)*crossentropyCauchy(l0,s0,l1,s1,theta2)+theta1*crossentropyCauchy(l1,s1,l0,s0,1-theta2)-hmixCauchy(l0,s0,l1,s1,theta1);
\end{lstlisting}

To get the Bregman generator, we do:
\begin{lstlisting}[breaklines,backgroundcolor = \color{lightgray},basicstyle=\small]
expand(-hmixCauchy(l0,s0,l1,s1,theta));
string(%);
\end{lstlisting}

Then for example, we can calculate and export in \TeX{}:
\begin{lstlisting}[breaklines,backgroundcolor = \color{lightgray},basicstyle=\small]
F:-hmixCauchy(0,1,1,1,theta);
g:derivative(F,theta,2); 
ratsimp(%);
tex(%);
\end{lstlisting}
and plot the potential function:
\begin{lstlisting}[breaklines,backgroundcolor = \color{lightgray},basicstyle=\small]
plot2d(F,[theta,0,1]);
\end{lstlisting}

To obtain a closed-form formula for the Kullback-Leibler divergence between $m_{\theta_1}$ and $m_{\theta_2}$, we do:
\begin{lstlisting}[breaklines,backgroundcolor = \color{lightgray},basicstyle=\small]
expand(KLmix(l0,s0,l1,s1,theta1,theta2));
string(%);
\end{lstlisting}

We can check by symbolic/numeric calculations that the Kullback-Leibler divergence between $m_{\theta_1}$ and $m_{\theta_2}$ are symmetric:
\begin{lstlisting}[breaklines,backgroundcolor = \color{lightgray},basicstyle=\small]
DiffKL(l0,s0,l1,s1,theta1,theta2):=KLmix(l0,s0,l1,s1,theta1,theta2)-KLmix(l0,s0,l1,s1,theta2,theta1);
float(DiffKL(0,1,1,1,0.2,0.8));
\end{lstlisting}

To implement the case of $(l_0,s_0)=(0,1)$ and $(l_1,s_1)=(1,1)$ discussed in the main body, we execute the following {\sc Maxima} code:

\begin{lstlisting}[breaklines,backgroundcolor = \color{lightgray},basicstyle=\small]
F(theta):=theta*log((2*sqrt(1+theta-(theta*theta))+theta+2)/(2*sqrt(1+theta-(theta*theta))-theta+3)) + log((2*sqrt(1+theta-(theta*theta))-theta+3)/(20*%pi));
gradF(theta):=log((2*sqrt(1+theta-(theta*theta))+theta+2)/(2*sqrt(1+theta-(theta*theta))-theta+3));
BD(theta1,theta2):=F(theta1)-F(theta2)-(theta1-theta2)*gradF(theta2);
expand(F(theta1)-F(theta2)-(theta1-theta2)*gradF(theta2));
theta(eta):=(5*exp(2*eta) + 2*sqrt(5)*sqrt(exp(3*eta) - 2*exp(2*eta)+exp(eta)) - 3*exp(eta))/(5*exp(2*eta)-6*exp(eta)+5);
Fdual(eta):=theta(eta)*eta-F(theta(eta));
expand(theta(eta)*eta-F(theta(eta)));
\end{lstlisting}
This yields a closed-formula for $F^*_{0,1,1,1}(\eta)$ (albeit a long formula).

\bibliographystyle{plain}
\bibliography{BregmanManifoldMixture2CauchyBIB}

\end{document}